\documentstyle[sprocl]{article}

\begin{document}
\renewcommand\floatpagefraction{.9}
\renewcommand\topfraction{1}
\renewcommand\bottomfraction{1}
\renewcommand\floatsep{12pt}
\renewcommand\textfraction{0}
\renewcommand\intextsep{12pt}
\input psfig

\title{STATUS OF SOLAR MODELS\ \footnote{This talk 
is based upon 
continuing collaborative research of John Bahcall and 
M. H. Pinsonneault. The first stages of
this work were described at the symposium on The Inconstant Sun,
Naples, Italy, March 18, 1996, to be published in Memorie della
Societa, eds. G. Cauzzi and C. Marmolino.  
This talk will appear in Neutrino 96, Proceedings of the 17 
International Conference on Neutrino Physics and Astrophysics,
Helsinki, Finland, ed. K. Huitu, K. Enqvist, and J. Maalampi
(World Scientific, Singapore, 1996).
For both conferences, 
the talks
were given by Bahcall. Further information about solar neutrinos is
available at
http://www.sns.ias.edu/\raise2pt\hbox{${\scriptstyle\sim}$}jnb .} } 
\author{JOHN BAHCALL}
\address{Institute for Advanced Study, Princeton, NJ 08540}
\author{M. H. PINSONNEAULT}
\address{Department of Astronomy, Ohio State University, Columbus, OH 
43210}

\maketitle\abstracts{
The neutrino fluxes calculated from 
14 standard solar models published recently in refereed
journals are inconsistent with the results of the 4
pioneering solar neutrino experiments if nothing happens to the
neutrinos after they are created in the solar interior. The sound
speeds calculated from 
standard solar models are in excellent agreement with
helioseismological measurements of sound speeds.
Some statements made by Dar at Neutrino 96 are answered
here. }

\section{Introduction}
\label{intro}

I was asked by Matts Roos to review in this talk the status of solar
models as they relate to the solar neutrino problems.  I will
therefore not discuss any of the solutions that suggest new physics;
that subject has just been covered beautifully by Alexei Smirnov and
there will be a further careful discussion by S. Petcov this afternoon.
I do, however, want to make a few introductory remarks in order to 
put my talk in the appropriate context.

Solar neutrino research has achieved its primary goal, the 
detection of solar neutrinos, and is now entering a new phase in which
large electronic detectors will yield vast amounts of diagnostic data.
The new experiments,~\cite{Arp92,Takita93,McD94} which will be
described after lunch today in talks by Suzuki, McDonald, Bellotti,
Vogelaar, and Bowles,   will test the prediction of standard
electroweak theory~\cite{Glashow61,Wein67,Salam68} 
that essentially nothing happens to electron type
neutrinos after they are created by nuclear fusion reactions in the
interior of the sun.

The four pioneering experiments---chlorine~\cite{Davis64,Davis94} 
(reviewed by Ken Lande at this conference), 
Kamiokande~\cite{Suzuki95} (reviewed by Y. Suzuki),  
GALLEX~\cite{Ansel95} (reviewed by T. Kirsten), and SAGE~\cite{Abdur94} 
(reviewed by V. Gavrin)---have 
all observed neutrino fluxes with intensities that are within a
factors of a few of  
those predicted by standard solar models. 
Three of the experiments (chlorine, GALLEX, and SAGE) are
radiochemical and each radiochemical experiment  measures
 one number, the total rate at which
neutrinos above a fixed energy threshold (which depends upon the
detector)  are captured.  
The sole electronic (non-radiochemical) 
detector among the initial experiments,
Kamiokande, has shown 
that the neutrinos come from the sun,
by measuring the recoil directions of the  electrons scattered by
solar neutrinos.
Kamiokande has also demonstrated 
that the observed neutrino energies 
are consistent with 
 the range of energies expected on the basis of the standard solar model.

Despite continual refinement of solar model calculations of
neutrino fluxes over the past 35 years (see, e.g., the collection of
 articles reprinted in the book edited by 
Bahcall, Davis, Parker, Smirnov, and Ulrich~\cite{BDP95}),
the discrepancies between 
observations and calculations have gotten worse with time.  All four
of the pioneering solar neutrino experiments yield event rates that
are significantly less than predicted by standard solar models.
Moreover, there are well known inconsistencies between the different
experiments if the observations are interpreted assuming that nothing
happens to the neutrinos after they are created.

This talk is organized as follows.
I will first summarize the results of all the 
recently published
standard solar model calculations and compare them with the results of
the four solar neutrino experiments.  
This survey of the literature is, to the best of my knowledge,
complete until June 1, 1996, just prior to the beginning of the
Neutrino 96 conference.
Next I shall discuss the excellent agreement
between the sound speeds predicted by standard solar models and the
sound speeds measured by helioseismological techniques. Finally, I
shall discuss briefly some of the remarks about solar models that
were made at the conference by A. Dar.

\section{Observation versus Calculation: Neutrino Fluxes}
\label{comparison}

\begin{figure}[t]
\centerline{\psfig{figure=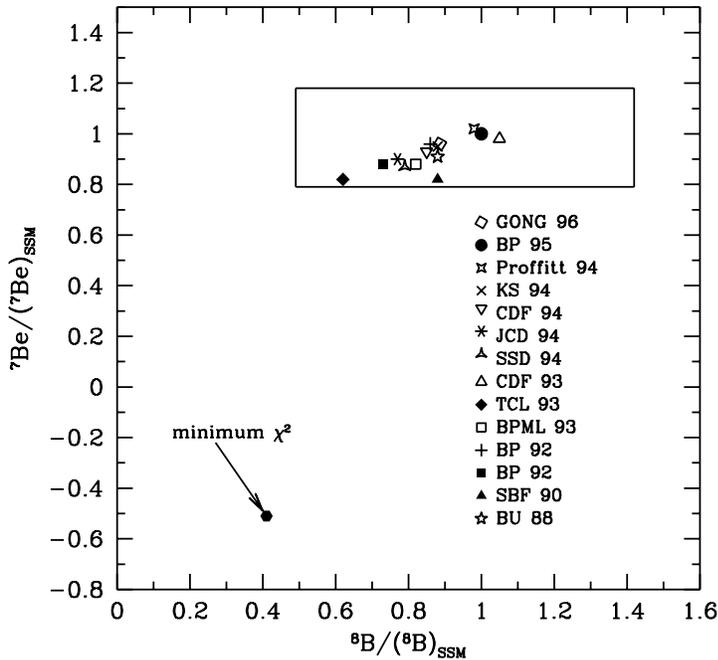,width=4in}}
\vglue-.2in
\caption[]{The calculated ${\rm ^7Be}$ and ${\rm
^8B}$ solar neutrino fluxes for all 14 of the standard solar models.
The sides of the rectangular box represent the estimated 
3$\sigma$ uncertainties in the predicted $^7$Be and $^8$B
neutrino fluxes of the standard solar model.~\cite{BP95}
All of the fluxes have been normalized by dividing by the Bahcall and
Pinsonneault~\cite{BP95} standard solar model (SSM) values. 
The abbreviations of the various solar models are GONG
(Christensen-Dalsgaard et al.~\cite{GONG}), BP
95 (Bahcall and Pinsonneault~\cite{BP95}),  KS 94 
(Kovetz and Shaviv~\cite{KS94}), CDF 94 (Castellani et al.~\cite{CDF94}), JCD 94
 (Christensen-Dalsgaard~\cite{JCD94}), SSD 94 (Shi, Schramm, and
Dearborn~\cite{SSD94}), CDF 93 (Castellani, Degl'Innocenti, and
Fiorentini~\cite{CDF93}), TCL 93 (Turck-Chi\`eze and Lopes~\cite{TCL93}), BPML 93 (Berthomieu, Provost,
Morel, and Lebreton~\cite{BPML93}), BP 92 
(Bahcall and Pinsonneault~\cite{BP92}), 
SBF 90 (Sackman, Boothroyd, and Fowler~\cite{SBF90}), and 
BU 88 (Bahcall and Ulrich~\cite{BU88}).}
\label{fluxes}
\end{figure}

Figure~\ref{fluxes} displays the calculated ${\rm ^7Be}$ and ${\rm
^8B}$ solar neutrino fluxes for all 14 of the standard solar models
with which I am  familiar that have been published in refereed
science journals since 1988 and
until June 1, 1996.  
I choose to start in 1988 since, as we shall see below, helioseismology
plays an important role in validating and constraining solar models
and the first systematic discussion of the
relation between helioseismology and solar neutrino research was
published in 1988.~\cite{BU88} I normalize the fluxes by dividing each
published value by the flux from the most recent 
Bahcall and Pinsonneault~\cite{BP95}
standard
solar model which makes use of improved 
input parameters and
includes heavy element and helium diffusion.  The abscissa is the
normalized ${\rm ^8B}$ flux and the numerator
is the normalized ${\rm ^7Be}$
neutrino flux.  
The sides of the rectangular box represent the separate
3$\sigma$ uncertainties in the predicted $^7$Be and $^8$B
neutrino fluxes of the standard solar model.~\cite{BP95}
The abbreviations that indicate references to individual models are
identified in the caption of Figure~\ref{fluxes}. 

All of the solar model results from different groups 
fall within the rectangular error box, i.e., within the 
estimated 3$\sigma$
uncertainties in the standard model predictions.  
This agreement between the results of 14 groups 
demonstrates the robustness of the predictions since the
 calculations use different computer codes and 
involve a variety of choices for the nuclear
parameters, the equation of state, the stellar radiative opacity, 
the initial heavy element abundances, and the physical processes
that are included.
In fact, all published standard  solar
models give the same results for solar neutrino fluxes to an accuracy
of better than 10\% if the same input parameters and physical
processes are included.~\cite{BP92,BP95}

The largest contribution to  the dispersion in values in
Figure~\ref{fluxes} is
caused by  the inclusion, or non-inclusion, of element diffusion in the 
stellar evolution codes.
The Proffitt,~\cite{Prof94} the
Bahcall and Pinsonneault,~\cite{BP95} and the 
Christensen-Dalsgaard et al.~\cite{GONG}  models all  
include helium and heavy element
diffusion.  The 
predicted fluxes in these three models agree to within $\pm 10\%$,
although the  models
are calculated  using different mathematical descriptions of
diffusion (and somewhat different input parameters),
The calculated value  that is furtherest from the center of the box 
is by Turck-Chi\`eze and Lopes,~\cite{TCL93} which does not include
either helium or heavy element diffusion.  However, the Turck-Chi\`eze
and Lopes best estimate is still well within the 3$\sigma$ box.

We shall now  see that helioseismology shows that diffusion must be
included in the solar model in order to obtain agreement with
observations.

\section{Comparison with Helioseismological Measurements}
\label{helioseismology} 

Helioseismology has recently sharpened
the disagreement between observations and the predictions of 
solar models 
with standard (non-oscillating) neutrinos.
The solar models that include
diffusion predict~\cite{BP95}  somewhat higher 
event rates in the chlorine and
Kamiokande solar neutrino experiments and thereby 
exacerbate the well known solar
neutrino problems that arise when standard neutrino physics (no
neutrino 
oscillations) is assumed.

By including element diffusion, 
the four solar models  near the center of the box in
Figure~\ref{fluxes} (models of Bahcall and Pinsonneault,~\cite{BP92} 
Proffitt,~\cite{Prof94} Bahcall and Pinsonneault,~\cite{BP95} 
and Christensen-Dalsgaard et al.~\cite{GONG})
yield values for the depth of the convective zone  and the primordial
helium abundance that are in agreement with helioseismological
measurements. 
(The model of Richard et al.~\cite{Rich96} yields results in good
agreement with the four solar models just mentioned that include
element diffusion, but was not yet published in Astron. and Astrophys.
by the cutoff date, June 1, 1996.)

\begin{figure}[t]
\centerline{\psfig{figure=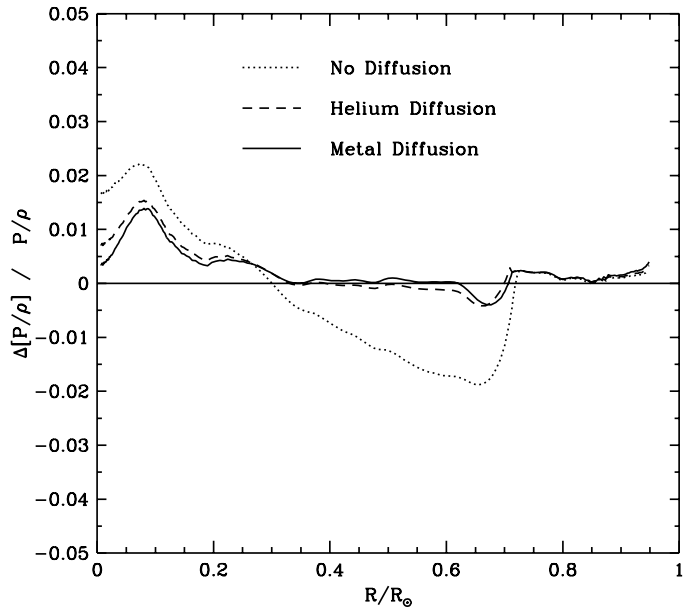,width=3.5truein}}
\caption[]{Comparison of  the profile of (pressure/density)
  predicted by different
standard solar models with the
values inferred from   helioseismology.  
There are no free parameters in the models; the microphysics is
successively improved by first including helium diffusion and 
then by using
helium and heavy element diffusion.
The figure shows the
fractional difference, ${[x -x\odot]}/{x_\odot} $, 
between the predicted Model values of $x = P/\rho$ and the 
measured~\cite{}
Solar values of $P/\rho$,  as a function of radial position in the sun
($R_\odot$ is the solar radius).
The dotted line refers to a model~\cite{BP95} in which
diffusion is not included  and the dashed  line was computed
from a model~\cite{BP95} in which helium  diffusion was included.
The dark line represents our   best 1995 solar model which includes 
both helium and heavy element diffusion.
\label{helio}}
\end{figure}

Figure~\ref{helio} compares the  values
of $P/\rho$ (pressure divided by density) obtained from
helioseismology and the values calculated for three different solar
models. The helioseismological values were kindly supplied to us by 
W. A. Dziembowski; they are based upon the Dziembowski et al. (1994)
method.~\cite{Dziembowski94}  The specific calculations leading to these improved values of
$P/\rho$ are described in Richard et al. (1996).~\cite{Rich96}  
The calculations
make use of new data for the low degree modes, $l \leq 3$, from the
BISON network.~\cite{Elsworth94}  

For the models that include  helium 
diffusion or helium plus heavy element diffusion, 
the agreement is excellent
between model predictions and the solar values of $P/\rho$.  
Over the entire region of the sun for which the helioseismological
values are well determined, from $0.3 \leq (r/R_\odot) \leq 0.95$, the
model values of $P/\rho$ agree with the helioseismological values to 
much better than 1\%.  To a good approximation, $P/\rho \propto
T/\mu$, where $T$ is the local value of the temperature and $\mu$ is
the local mean molecular weight.  The temperature in the standard solar
model changes by a factor of
$24$ from $R = 0.3 R_\odot$ to $R = 0.95 R_\odot$, while the molecular
weight only changes by a few percent. 

The excellent agreement shown in Figure~\ref{helio}
between solar models that include  diffusion and the
helioseismological observations 
demonstrates that solar models correctly predict 
the temperature profile of the sun to a few tenths of a
percent over most of the sun.
The agreement is less precise, of the order of 1\%, in the deep
interior, but in this region the observations are not yet very
reliable.

Helioseismology, as summarized in Figure~\ref{helio}, 
 has effectively shown that the solar neutrino problems
cannot be ascribed to errors in the temperature profile of the sun. It
is well known~\cite{BU96,Castellani94} 
that in order to change the predicted neutrino fluxes by
amounts sufficient to affect significantly 
the discrepancies with neutrino observations
the temperatures must differ from the values in the standard solar
model by at least 5\%.
Figure~\ref{helio} shows that helioseismology constrains the
differences from standard models to be everywhere less than or of
order 1\%, and much less than 1\% over most of the sun.

Solar models that do not include diffusion 
are not consistent with the helioseismological evidence (for previous
evidence supporting this conclusion see the  discussions in
Christensen-Dalsgaard, Proffitt, and Thompson,~\cite{CDPT93} 
Guzik and Cox,~\cite{GC93} Bahcall and Pinsonneault,~\cite{BP95} and
Christensen-Dalsgaard et al.~\cite{GONG}).  Figure~\ref{helio} shows
that solar models in which diffusion is not included are grossly 
inconsistent
with the helioseismological observations in the region in which the 
observations are most reliable and precise.

In my view, only solar models that include element diffusion should,
in the future,  be
called ``standard solar models''. 
These ``standard models'' all lie close to the center of the
rectangular error box in Figure~\ref{fluxes}.
 The physics of
diffusion is simple and there is an exportable subroutine available
for calculating diffusion in stars (see
http://www.sns.ias.edu/$^\sim$jnb).  Observation requires, and
computing technology 
easily permits, the inclusion of diffusion in any standard
stellar evolution code.

\section{Recent Improvements in the Equation of State and Opacity}
\label{improvements}

In preparation for this meeting, we have calculated new solar models
that include recent improvements in
opacity~\cite{Opacity} and equation of state~\cite{eos} on the predicted
solar neutrino fluxes.  Table~\ref{tablefluxes} gives the neutrino fluxes
computed for 
three different standard solar models, all of which include helium
and heavy element diffusion.  The model labeled BP95 is
from Bahcall and Pinsonneault;~\cite{BP95} the models labeled New Opac and OPAL
 EOS include,
respectively, the improved opacities discussed in 
Iglesias and Rogers~\cite{Opacity} and
the improved opacities plus the new OPAL equation of state 
discussed in Rogers, Swenson, and Iglesias.~\cite{eos}

\begin{table}[htb]
\centering
\caption[]{Neutrino Fluxes for Solar Models with Diffusion.  All
fluxes, except for $^8$B and $^{17}$F, 
 are given in units of $10^{10}$ per ${\rm cm^{-2}s^{-1}}$ at
the earth's surface. The $^8$B and $^{17}$F fluxes are in units of 
$10^{6}$ per ${\rm cm^{-2}s^{-1}}$.}
\vspace{0.4cm}
\begin{tabular}{|lccccccc|}
\hline
&&&&&&&\\[-8pt]
\hfil Model\hfil&$pp$&$pep$&${\rm ^7Be}$&${\rm ^8B}$&${\rm
^{13}N}$&${\rm ^{15}O}$&${\rm ^{17}F}$\\
&&&&&&&\\ [-8pt]\hline 
&&&&&&&\\ [-8pt]
BP95&5.91&0.014&0.515&0.662&0.062&0.055&0.065\\
New Opac&5.91&0.014&0.516&0.662&0.062&0.055&0.065\\
OPAL EOS&5.91&0.014&0.513&0.660&0.062&0.054&0.065\\
\hline
\end{tabular}
\label{tablefluxes}
\end{table}

The neutrino fluxes computed with the improved opacity and equation of
state  differ from the previously published values~\cite{BP95} by 
amounts that are negligible in solar neutrino calculations.
The predicted event rate, for all three models, 
is 
\begin{equation}
{\rm Cl~Rate} ~=~9.5^{+1.2}_{-1.4}~~{\rm SNU}
\label{clrate}
\end{equation}
for the chlorine experiment 
and 
\begin{equation}
{\rm Ga~Rate} ~=~137^{+8}_{-7}~~{\rm SNU}
\label{garate}
\end{equation}
for the gallium experiments.
The only noticeable change in the predicted
event rates for the chlorine and the gallium experiment is a slightly
increased (by 2\%) event rate for chlorine, which is due to a small
improvement~\cite{b8cross} 
in the calculation of the neutrino absorption cross
sections for $^8$B.  

It is obviously important to compare the improved solar models with
helioseismological measurements to see if the better equation of state
and opacity used in these most recent models affect significantly the
calculated sound velocities.  Unfortunately, we were not able to
complete those calculations in time for the meeting.

\section{Quantitative Comparison with Neutrino Experiments}
\label{experimentalcomparison}

How do the 
observations from the four pioneering solar neutrino
experiments agree with the solar model calculation?  
Plamen Krastev and I (see Bahcall and Krastev~\cite{BK96} for a
description of the techniques) have recently compared the
predicted standard model fluxes, with their estimated uncertainties,
and the observed rates in the chlorine, Kamiokande, GALLEX, and SAGE
experiments.  The theoretical solar model and experimental
uncertainties, as well as the uncertainties in the neutrino cross
sections, have been
combined quadratically. Using the predicted fluxes from the Bahcall
and Pinsonneault~\cite{BP95}
model, the $\chi^2$ for the fit to the four experiments is

\begin{equation}
\chi^2_{\rm SSM} \hbox{(all 4 experiments)} = 56\ .
\label{chitofit}
\end{equation}
The theoretical uncertainties (from the solar model and the neutrino
cross section calculations) and the experimental errors (statistical
and systematic, combined quadratically) have been taken into account
 in obtaining Eq. 
\ref{chitofit}.

Suppose we now ignore what we  have learned  from solar models and allow
the important 
${\rm ^7Be}$ and ${\rm ^8B}$ fluxes to take on any non-negative
values.  What is the minimum value of $\chi^2$ for the 4 experiments,
when the only constraint on the fluxes is the requirement that the
luminosity of the sun be supplied by nuclear fusion reactions among
light elements?  We include the nuclear physics inequalities between
neutrino fluxes (see section 4 of Bahcall and Krastev~\cite{BK96}) that are
associated with the luminosity constraint and maintain the standard
value for the almost
model-independent  ratio of $pep$ to $pp$ neutrinos.

The best fit for arbitrary $^7$Be and $^8$B neutrino fluxes 
is obtained for ${\rm ^7Be/(^7Be)}_{\rm SSM} = 0$ and
${\rm ^8B/(^8B)}_{\rm SSM} = 0.40$, where

\begin{equation}
\chi^2_{\rm minimum} \hbox{(all 4
experiments; ~arbitrary $^7$Be, $^8$B)} = 14.4\ . 
\label{chiall}
\end{equation}
The CNO neutrinos were assumed equal to their standard model values in
the calculations that led to Eq.~\ref{chiall}.  The fit can be further
improved if we set the CNO neutrino fluxes equal to zero. 
Then, the same search for arbitrary $^7$Be and
$^8$B neutrino fluxes leads to 

\begin{equation}
\chi^2_{\rm minimum} \hbox{(all 4
experiments; ~arbitrary $^7$Be, $^8$B; CNO = 0)} = 5.9\ .
\label{chinoCNO}
\end{equation}

If we drop the physical requirement that the $^7$Be flux be positive
definite, the minimum $\chi^2$ occurs (cf. 
Figure~\ref{fluxes}) for a negative value of the ${\rm ^7Be}$
flux; this unphysical 
result is a reflection of what has become  known in the physics
literature as `` the  missing ${\rm
^7Be}$ solar neutrinos.''.  The reason that the $^7$Be neutrinos appear
to be missing (or have a negative flux) is that the two gallium
experiments, GALLEX and SAGE, have an average event rate of $74 \pm 8$
SNU,
which is fully accounted for in the standard model 
by the fundamental $p-p$ and $pep$ 
neutrinos (best estimate $73 \pm 1$ SNU). 
In addition, the $^8$B neutrinos
that are observed in the Kamiokande experiment will produce about $7$
SNU in the gallium experiments, unless new particle physics affects
the neutrinos.

To me, these results suggest strongly that the assumption on
which they are based---nothing happens to the neutrinos after they are
created in the interior of the sun---is incorrect.  A less plausible 
 alternative
(in my view)  is that some of the
experiments are wrong; this must be checked by further experiments.

\section{Comments on Some Remarks by Dar}
\label{dar}

In the closing session on  solar neutrinos
at Neutrino~96, Arnon Dar made a   number of 
surprising statements
about solar models and the input data
used in their construction.~\cite{Dar96}
I  state  below in italics some of  Dar's  most  remarkable 
claims.  The resolution of each of the issues he raised is
given in a  paragraph following the relevant italicized statement.

\bigskip
{$\bullet$ \it  Final state interactions in $^{37}$Cl and $^{71}$Ga may
invalidate the neutrino cross sections of Bahcall for low energy
$pp$ and $^7$Be neutrinos.}

Dar cites electron screening,  overlap and exchange effects, nuclear
recoil, and radiative corrections  as final
state interactions that might be important.

Electron screening is included explicitly in Bahcall's calculations 
with the aid of Hartree-Fock
wave functions and amounts to an effect of order 1\% for $^{37}$Cl and
4\% for $^{71}$Ga. 
Overlap and exchange
effects, as well as bound-state beta-decay, were evaluated in
Section~III of Bahcall (1978) and found to be less than 1\%. These
results are summarized in Section~8.1A of the book {\it 
Neutrino Astrophysics}.~\cite{Bahcall89} 
Radiative corrections have
been calculated explicitly for some cases and are
about $1$\% (i.e, of order the fine structure constant, $\alpha$).
Nuclear recoil effects are
$\sim$ [nuclear recoil energy/(electron kinetic energy)] and are less
than $0.1$\% for $^{37}$Cl and $^{71}$Ga. 

\medskip
{$\bullet$ \it  A strong magnetic field may polarize the electrons in the
solar interior and affect the branching ratios of electron capture by 
$^7$Be.}

In order to polarize electrons in the solar
interior with typical kinetic energies of order a keV, 
a magnetic field of order $10^{12}$G
is required.  
A field of $10^{12}$G would produce a total pressure in the solar
interior $10^5$ times larger than the pressure in standard solar
models and is therefore ruled out by the excellent agreement 
(to within $1$\%) between
the standard models and the helioseismological measurements
(see Figure~\ref{helio} and Section~5.6 of {\it Neutrino Astrophysics}).

\medskip
{$\bullet$ \it Something must be wrong because it is known that the 
 OPAL equation of state causes significant changes in
the calculated neutrino fluxes.}

In his talk, Dar cited calculations in which the use of the OPAL
equation of state significantly affected the calculated neutrino
fluxes.  
He suggested that something must be wrong with our calculations
because we did not find large changes when we used the new equation of
state.

The previous equation of state and opacity values 
that we have been using are quite
close in the solar interior to the  newer OPAL equation
of state and opacity tables.
This explains why we find only
small changes in  the neutrino fluxes 
(see Table~\ref{tablefluxes}).
Presumably, for the codes Dar cited, the new OPAL data caused
larger changes in the input physics and hence larger changes in the
calculated neutrino fluxes.

\medskip
{$\bullet$ \it The differences between the Bahcall-Pinsonneault and the
Dar-Shaviv nuclear reaction cross sections represent personal
judgment.}

We use the cross section factors published by the experimentalists who
did the measurements.  When multiple measurements are made of a given
reaction, we use the weighted average of the measurements that is
published by nuclear physicists. 

Dar described in his talk his proposed  method  of extrapolation,
which is apparently 
different from what nuclear experimentalists  have
traditionally used.  
The analysis by Dar has been criticized by Langanke,~\cite{Langanke95}
 who argues that a
proper treatment with Dar's method must lead to the same results as
obtained by the more traditional extrapolation.
Dar and Shaviv use  six  cross section factors
that are significantly different from the conventional values that we
have taken from the literature.~\cite{Dar96}  
All of the choices that Dar and
Shaviv have made are in the direction of reducing the calculated event
rates in the solar neutrino experiments.

\medskip
{\it $\bullet$  
The $pp$ reaction cross section can be calculated accurately 
from measured
reactions involving deuterium.}

As justification for his choice of 
the cross section factor  for  the
$^1H(p,e^+ + \nu){^2}$H reaction,  Dar cites~\cite{Dar96} the 
experimental cross sections for anti-neutrinos and gamma-rays 
 on deuterium.  He    states that these measurements 
were used to obtain a
cross section for $p + p \rightarrow  {^2H} + e^+ +\nu$.  No 
equations or other details
are given (see page 938 of ref.~37).
The most relevant measurement to which Dar refers is the reaction
 ${\bar \nu_e} + {^2H} \rightarrow  n + n +e^+$, for which the quoted
$1\sigma$ experimental uncertainty is $26$\%.~\cite{Pasierb79}
The matrix element 
for the $\gamma$-disintegration reaction he cites is not
the same as the matrix element for the neutrino reaction.
Dar states that his procedure yields a value consistent
with the Caughlan and Fowler (1988) rate,~\cite{Fowler} 
which was based upon the recalculation of the $pp$ cross section factor 
by Bahcall and Ulrich (1988).~\cite{BU88}  
Since the publication of
the 1988 work, Kamionkowski and Bahcall(1994)~\cite{KB94} 
included vacuum polarization in the
calculation of this reaction and reevaluated the nuclear matrix
elements using improved data for the $pp$ scattering and for the
deuteron wave function.  In their published paper, Kamionkowski and
Bahcall tabulated the numerical results they obtained by solving the
Schroedinger equation with seven different nuclear potentials that have
been used by different nuclear physics groups.  Combining the
theoretical and experimental uncertainties, Kamionkowski and Bahcall
find $S_{pp}(0) = 3.89(1 \pm 0.011)~{\rm MeV~barns}$.  Dar gives a
value of $\approx 4.07$, with no quoted uncertainty, 
 instead of $3.89(1 \pm 0.011)$.
The neutrino cross sections to which Dar refers as his justification
are uncertain by much more than the $4$\% difference
between his estimated value and 
the detailed  Kamionkowski and Bahcall calculation.
From the published literature, one cannot determine how Dar obtained
the value he quotes.

\medskip
{$\bullet$ \it The $pep$ and $^7$Be electron capture rates of Bahcall and
his collaborators are not as accurate as 
the 1988 tabulation by Fowler and Caughlan
.}

This statement is based upon a misunderstanding of the purpose of the
Fowler-Caughlan tabulations.

The Fowler and Caughlan expressions are simple 
analytic approximations to the complicated 
expressions derived by  Bahcall and his collaborators.  The Fowler and
Caughlan expressions are designed to be approximately valid, as they
state, over an enormous range of temperatures, $10^6$~K to $10^9$~K.
They are not designed to reproduce precisely, for solar temperatures,
the expressions of Bahcall et al. from which they are derived.

For both the $pep$ and the $^7$Be electron capture 
 reactions, all of the references by Fowler, Caughlan, and
their collaborators in their 5 review articles~\cite{Fowler} 
are to results by
Bahcall and his collaborators (see ref. [41]).

\medskip
{$\bullet$ \it The normalization of the heavy element abundances is not
handled properly by Bahcall and Pinsonneault.}

Dar states that Bahcall and Pinsonneault assume that the ``present 
photospheric abundances equal the meteoritic abundances.''

This is not only an incorrect statement of what we do, it is
impossible to implement.  Meteorites are rocks; they do not contain
hydrogen. Therefore, one cannot fix the normalization of the heavy
elements from the meteoritic abundances.

As described at the bottom of page 87 of {\it Neutrino Astrophysics}, 
we take the {\it relative} abundances
of the heavy elements (except for He,C,N,O, and Ne) from
the meteorites.  We assume that this set of relative abundances
applies to the {\it initial} sun.  
The connection between  the meteoritic abundances and the 
measured solar
photospheric abundances, which do include hydrogen, is made 
by Anders and Grevesse~\cite{Anders89} using a
series  of elements for which abundances are measured accurately in
both the photosphere and the meteorites.  This fixes $Z/X$ on the
surface of the sun today, where $X$ and $Z$ are the mass fractions of
hydrogen and heavy elements.  Of course, $X$ + $Y$ + $Z$ = 1, where
$Y$ is the helium mass fraction.
We fix the absolute values of the 
abundances by requiring that the current solar
model have a luminosity at the present solar epoch equal to the
observed solar luminosity.  
With the normalization of the three fractions, the observed ratio of 
$Z/X$, and the luminosity constraint, we have three equations for
three unknowns.
In models in which diffusion is included,
the current surface  abundance of heavy elements 
is different from the initial surface abundance of heavy elements.

It is not clear  how Dar and Shaviv normalize their
heavy element abundances since Dar states that they assume that the
``initial solar abundance equals the meteoritic abundance.''
As explained above, the meteorites only determine relative 
heavy element abundances.

\section{Discussion}
\label{closing}

The combined predictions of the standard solar model and the standard
electroweak theory disagree with the results of the four pioneering
solar neutrino experiments.  
The same solar model calculations are in good agreement with the
helioseismological measurements.

Comparing the solar model  predictions to the existing solar neutrino 
data, we obtain
  values for
$\chi^2_{standard}$ of  $\sim 56$.
The fits are
much improved if neutrino oscillations, which are described by  two free
parameters, are included in the calculations.  With neutrino
oscillations, the characteristic value for $\chi^2_{\min, ~osc.} \sim
1$.  New experiments~\cite{Arp92,Takita93,McD94} 
involving large electronic detectors of
individual neutrino events will decide in the next few years if
neutrino oscillations are indeed important in interpreting solar
neutrino experiments.

We may ask:  What have solar neutrino experiments taught us
about astronomy? 
Most importantly, the  experiments have detected solar
neutrinos with approximately the fluxes and in the energy range
predicted by solar models.
The operating 
experiments have achieved the initial goal of
solar neutrino astronomy by showing empirically that the sun shines
via nuclear fusion reactions.  
This achievement by a large community of physicists, chemists,
engineers, and astronomers puts the theory of stellar evolution on a
firm empirical basis.

Moreover, the observed and the standard
predicted
neutrino interaction rates agree within  factors of a few, providing---
even if we ignore the effects of possible neutrino oscillations---
semi-quantitative confirmation of the calculations of
temperature-sensitive nuclear fusion rates
in the solar interior.

\section*{Acknowledgments}
We are  grateful to Plamen Krastev for valuable conversations and for
performing the $\chi^2$ calculations.
The research of John Bahcall is supported in part by NSF
grant number PHY95-13835.


\section*{References}
\begin{thebibliography}{99}
\bibitem{Arp92}C. Arpesella et al., BOREXINO proposal, Vols. 1
and 2, eds. G. Bellini, R. Raghavan, et al. (Univ. of Milano, 1992).
\bibitem{Takita93}M. Takita, in {\it Frontiers of Neutrino
Astrophysics\/}, eds. Y. Suzuki and K. Nakamura (Universal Academy 
Press, 1993) p. 147.
\bibitem{McD94}A.B. McDonald, in {\em Proceedings of the 9th Lake 
Louise Winter Institute\/}, eds. A.\ Astbury et al. (World Scientific, 1994)
p. 1.
\bibitem{Glashow61}S.L. Glashow, {\em Nucl. Phys.\/} {\bf 22}, 579 (1961).
\bibitem{Wein67}S. Weinberg, {\em Phys. Rev. Lett.\/} {\bf 19}, 1264 (1967).
\bibitem{Salam68}A. Salam, in {\em Elementary Particle Theory\/}, ed.  
N. Svartholm (Almqvist and Wiksells, 1968) p. 367.
\bibitem{Davis64}R. Davis, Jr., {\em Phys. Rev. Lett.\/} {\bf 12}, 303 (1964).
\bibitem{Davis94}R. Davis, Jr., {\em Prog. Part. Nucl. Phys.\/} {\bf
32}, 13 (1994).
\bibitem{Suzuki95}Y. Suzuki, KAMIOKANDE collaboration, {\em Nucl. Phys. B 
(Proc. Suppl.)\/} {\bf 38}, 54 (1995).
\bibitem{Ansel95}P. Anselmann, et al., GALLEX collaboration, 
{\em Phys. Lett. B\/} {\bf 342}, 440 (1995).
\bibitem{Abdur94}J.N. Abdurashitov, et al., SAGE Collaboration,   
{\em Phys. Lett. B\/} {\bf 328}, 234 (1994).
\bibitem{BDP95}Eds. J.N. Bahcall, R. Davis, Jr., P. Parker, A. Smirnov, and
R.K. Ulrich, {\em Solar Neutrinos: The First Thirty
Years\/} (Addison Wesley 1995).
\bibitem{BU88}J.N. Bahcall and R.K. Ulrich, {\em Rev. Mod. Phys.\/}
{\bf 60}, 297 (1988).
\bibitem{BP95}J.N. Bahcall, M.H. Pinsonneault, 
{\em Rev. Mod. Phys.\/} {\bf 67}, 781 (1995).
\bibitem{GONG}J. Christensen-Dalsgaard et al., GONG collaboration, 
{\em Science\/} {\bf 272}, 1286 (1996).
\bibitem{KS94}A. Kovetz and G. Shaviv, {\em Astrophys. J.\/} {\bf
426}, 787 (1994).
\bibitem{CDF94}V. Castellani, S. Degl'Innocenti, G. Fiorentini, L.M.
Lissia, and B. Ricci, {\em Phys. Lett. B\/} {\bf 324}, 425 (1994).
\bibitem{JCD94}J. Christensen-Dalsgaard, {\em  Europhysics News\/}
{\bf 25}, 71 (1994).
\bibitem{SSD94}X. Shi, D.N. Schramm, D.S.P. Dearborn, 
{\em Phys. Rev. D\/} {\bf 50}, 2414 (1994).
\bibitem{CDF93}V. Castellani, S. Degl'Innocenti, and G. Fiorentini, {\em
Astron. Astrophys\/} {\bf 271}, 601 (1993).
\bibitem{TCL93}S. Turck-Chi\`eze and I. Lopes, {\em Astrophys.
J.\/} {\bf 408}, 347 (1993).
\bibitem{BPML93}B. Berthomieu, J. Provost, P. Morel, and Y. Lebreton,
{\em Astron. Astrophys.\/} {\bf 268}, 775 (1993).
\bibitem{BP92}J.N. Bahcall and M.H. Pinsonneault, 
{\em Rev. Mod. Phys.\/} {\bf 64}, 885 (1992).
\bibitem{SBF90}I.-J. Sackman, A.I. Boothroyd, and W.A. Fowler, {\em
Astrophys. J.\/} {\bf 360}, 727 (1990).
\bibitem{Prof94}C.R. Proffitt, {\em Astrophys. J.\/} {\bf 425}, 849
(1994).
\bibitem{Rich96}O. Richard, S. Vauclair, C. Charbonnel, and W.A.
Dziembowski, 
{\em  Astron. Astrophys\/}, in press (1996).
\bibitem{Dziembowski94} W.A. Dziembowski, P.R. Goode, A.A.
Pamyatnikh, and R. Sienkiewicz, {\em Astrophys. J.\/} 
{\bf 432}, 417 (1994).
\bibitem{Elsworth94} Y. Elsworth, R. Howe, G.R. Issak, C.P. McLeod,
R. New, {\em Astrophys. J.\/} {\bf 434}, 801 (1994).
\bibitem{BU96}J.N. Bahcall and A. Ulmer, Phys. Rev. D {\bf 53}, 4202
(1996). 
\bibitem{Castellani94}V. Castellani, S. Degl'Innocenti, G. Fiorentini,
and M. Lissia, Phys. Rev. D. {\bf 50}, 4749 (1994).
\bibitem{CDPT93}J. Christensen-Dalsgaard, C.R. Proffitt, and 
M.J. Thompson, {\em Astrophys. J. Lett.\/} {\bf 403}, L75 (1993). 
\bibitem{GC93}A. Guzik and A.N. Cox, {\em Astrophys. J.\/} {\bf 411},
394 (1993).
\bibitem{Opacity}C.A. Iglesias and F.J. Rogers, Astrophys. J. {\bf 464}, 943
(1996); D.R. Alexander and J.W. Ferguson, Astrophys. J. {\bf 437},
879 (1994). The new OPAL opacities include more elements (19 rather than 12) and cover a wider range in temperature, density, and composition.  The low temperature opacity
tables include more opacity sources and a wider range of composition.
\bibitem{eos}F.J. Rogers, F.J. Swenson, and C.A. Iglesias,
Astrophys. J. {\bf 456},902 (1996).
\bibitem{b8cross}J.N. Bahcall et al. Phys. Rev. C {\bf
54}, 411 (1996).
\bibitem{BK96}J.N. Bahcall and P.I. Krastev, {\em Phys.\ Rev.\ D\/}
{\bf 53}, 4211 (1996).
\bibitem{Dar96}A. Dar and G. Shaviv, {\em Astrophys.~J.\/} {\bf 468},
933 (1996).  The material on which Dar based his talk was subsequently
published in this paper.
\bibitem{Bahcall89}J.N. Bahcall, {\em Neutrino Astrophysics\/} (Cambridge 
		University Press, 1989).
\bibitem{Langanke95}K. Langanke, in {\em Solar Modeling \/} ed.
A.B. Balantekin and J.N. Bahcall (World Scientific, London) (1995).
\bibitem{Pasierb79}S. Pasierb et al., {\em Phys. Rev. Lett.\/} {\bf
43}, 96 (1979). 
\bibitem{Fowler}G.R. Caughlan, W.A. Fowler, M.J. Harris, and B.A.
Zimmerman, {\em Atomic Data and Nuclear Data Tables} {\bf 40}, 283
(1988); {\em ibid.} {\bf 32}, 197 (1985); M.J. Harris, W.A. Fowler,
G.R. Caughlan, and B.A. Zimmerman, {\em Ann. Rev. Astron.
Astrophys.} {\bf 21}, 165 (1983); W.A. Fowler, G.R. Caughlan, and B.A.
Zimmerman, {\em Ann. Rev. Astron. Astrophys.} {\bf 13}, 69 (1975);
W.A. Fowler, G.R. Caughlan, and B.A. Zimmerman, {\em Ann. Rev.
Astron. Astrophys.} {\bf 5}, 525 (1967). 
In their first review, Fowler et al.
(1967) cited as the source of their quoted rate for the $^7$Be
reaction ``... Bahcall (1962)
who gives an approximate expression...(their equation 102).''
In their 1983 review, Fowler et al. state:  ``The reaction rates for
the proton-proton chain, which is of importance in the solar neutrino
problem, have been changed (with one exception) to correspond to the
analysis of experimental data made by Bahcall et al. (1982).  The
exception is the rate of the $^3He(\alpha, \gamma)^7Be$ reaction for
which we have included the recent results of Robertson ...''  The
final statement made in this connection 
by Fowler et al. was in their 1985 review in
which they state explicitly ``The coefficient in the rate for 
$^1H(p + e^-, \nu)^2H$ has been changed to correspond to the analyses
of Bahcall et al. (1982)...This change gives a result in better
agreement with the exact expression of Bahcall et al. (1982)...The
only change in $^7Be(e^-, \nu)^7$Li is the deletion of the $T9$ term
for the same reason that term has been deleted in the expression for 
$^1H(p + e^-, \nu)^2H$.'' 
\bibitem{KB94}M. Kamionkowski and J.N. Bahcall, {\em Astrophys. J.\/}
{\bf 420}, 884 (1994).
\bibitem{Anders89}E. Anders and N. Grevesse, Geochim. Cosmochim. Acta
{\bf 53}, 197 (1989).
\end{thebibliography}
\end{document}